%% file: main.tex
\documentclass[%
 reprint,
 superscriptaddress,
nofootinbib,
 amsmath,amssymb,
 aps,
 prl,
]{revtex4-2}

\usepackage{times}
\usepackage{newtxmath}

\usepackage{graphicx}
\usepackage{mathrsfs}
\usepackage{bm}
\usepackage{xcolor}
\usepackage{hyperref}
\usepackage{wasysym}
\usepackage{physics}
\usepackage{orcidlink}
\definecolor{darkblue}{rgb}{0.0, 0.1, 0.7}
\hypersetup{
    colorlinks=true,
    linkcolor=darkblue,
    filecolor=darkblue,      
    urlcolor=darkblue,
    citecolor = darkblue,
    pdftitle={Analytic Solution for the Motion of Spinning Particles in Kerr Space-Time},
    }

\usepackage{fancyhdr}
\begin{document}

  \newcommand{\Tt}{\Tilde{t}}
\renewcommand{\Tr}{\Tilde{r}}
  \newcommand{\Tz}{\Tilde{z}}
  \newcommand{\Tphi}{\Tilde{\phi}}
  \newcommand{\Tl}{\Tilde{\lambda}}

\newcommand{\mytitle}[1]{\noindent
\emph{#1.}--}

\newcommand{\vs}[1]{\textcolor{blue}{VS: #1}}
\newcommand{\vw}[1]{\textcolor{red}{VW: #1}}

\title{Analytic Solution for the Motion of Spinning Particles in Kerr Space-Time}

\author{Viktor Skoup\'y \orcidlink{0000-0001-7475-5324}}
\email{viktor.skoupy@matfyz.cuni.cz}

\author{Vojt{\v e}ch Witzany \orcidlink{0000-0002-9209-5355}}
\email{vojtech.witzany@matfyz.cuni.cz}

\affiliation{Institute of Theoretical Physics, Faculty of Mathematics and Physics, Charles University, CZ-180 00 Prague, Czech Republic}

\begin{abstract}
The equations of motion of massive test particles near Kerr black holes are separable in Boyer-Lindquist coordinates, as established by Carter. This separability, however, is lost when the particles are endowed with classical spin. We show that separability of the equations of motion can be recovered to linear order in spin by a shift of the worldline derived with the use of the hidden symmetry of Kerr space-time. Consequently, the closed-form solution of the motion is expressed in a way closely analogous to the solution for spinless particles. This finding enriches the understanding of separability and integrability properties of the dynamics of test particles and fields in Kerr space-time, and is particularly valuable for modeling inspirals of rotating compact objects into massive black holes.
\end{abstract}

\maketitle

%%%%%%%%%%%%%%%%%%%%%%%%%%%%%%%%%%%%%%%%%%%%%%%%%%%%%%%%%%%%%
\mytitle{Introduction}
Upcoming space-based gravitational-wave observatories such as LISA, TianQin, or Taiji \cite{LISA,LISA:2024,TianQin,Taiji} will detect signals from various sources, including intermediate- and extreme-mass-ratio inspirals of compact objects into massive black holes \cite{Babak:2017, LISAAstroWG:2022}. 

These detections will allow for testing general relativity in the strong fields near massive black holes \cite{Barack:2006pq}. To achieve this and other scientific goals \cite{LISAFunWG:2022,LISAAstroWG:2022}, we must generate gravitational-wave templates sourced by these inspirals. Black hole perturbation theory is typically used to generate waveforms from large mass ratio systems \cite{LISAWavWG:2023,Pound:2021} by modeling the lighter body's motion as corrected geodesic motion in the space-time of the massive black hole. Since astrophysical compact objects rotate at relativistic speeds, accurate waveform generation must account for both bodies' spins  \cite{LISAWavWG:2023}. The leading order spin effects are captured by the motion of a spinning particle in the space-time of a Kerr black hole.

Kerr space-time exhibits the so-called hidden symmetry, which is best expressed by the existence of the closed conformal Killing-Yano tensor (see Eq. \eqref{eq:fmunu}) \cite{Frolov:2017kze}. This property leads to the separation of the geodesic equation \cite{Carter:1968ks} and various field equations, including the Klein-Gordon, Dirac, and Teukolsky equations \cite{Chandrasekhar:1976ap,Teukolsky:1973ha}. Moreover, even for the trajectories of spinning particles in the linear-in-spin regime, it was proven that there exists a sufficient number of constants of motion to achieve integrability \cite{Rudiger:1981,Rudiger:1983,Dixon:1970I,Gibbons:1993ap,Kubiznak:2011ay}. Furthermore, the solution of the Hamilton-Jacobi equation for the spinning-particle Hamiltonian was constructed perturbatively, reducing the equations of motion to the first order \cite{Witzany:2018ahb,Witzany:2019}.

Nevertheless, the equations of motion were not shown to be separable and remained unsolved analytically for generic orbits until now. Analytical expressions exist only for fundamental frequencies of motion \cite{Witzany:2024ttz} while trajectory solutions have been obtained by numerically or semianalytically calculating Fourier coefficients of the nongeodesic corrections to the orbits \cite{Drummond:2022a,Drummond:2022b,Piovano:2024}. 

In this Letter, we introduce a coordinate shift of the spinning particle trajectory to a virtual worldline associated with a geodesic whose constants of motion are near those of the spinning particle. Consequently, the transformed equations of motion become separable, allowing the trajectory solution to be expressed using existing closed formulas for geodesic motion \cite{Fujita:2009,vandeMeent:2020,Cieslik:2023}.

%%%%%%%%%%%%%%%%%%%%%%%%%%%%%%%%%%%%%%%%%%%%%%%%%%%%%%%%%%%%%
\mytitle{Notation}
We use geometrized units with $G=c=1$ and the metric signature $(-,+,+,+)$. Space-time indices are denoted by lowercase Greek letters; spatial indices are denoted by lowercase Latin letters.
The Riemann tensor is defined as $a_{\nu;\kappa\lambda} - a_{\nu;\lambda\kappa} \equiv R^\mu{}_{\nu\kappa\lambda} a_\mu$, where the semicolon indicates the covariant derivative and $a_\mu$ is a covector. Round brackets around indices $T_{(\alpha\beta \ldots \gamma)}$ denote total symmetrization.

%%%%%%%%%%%%%%%%%%%%%%%%%%%%%%%%%%%%%%%%%%%%%%%%%%%%%%%%%%%%%
\mytitle{Motion of spinning bodies in curved space-time}
In the pole-dipole approximation, spinning bodies in curved space-time follow trajectories governed by the Mathisson-Papapetrou-Dixon (MPD) equations \cite{Mathisson:2010,Papapetrou:1951pa,Dixon:1970I}
\begin{align}\label{eq:MPD_full}
    \frac{D p^\mu}{\dd \tau} &= -\frac{1}{2} R^{\mu}{}_{\nu\rho\sigma} u^\nu S^{\rho\sigma} \,, & \frac{D S^{\mu\nu}}{\dd \tau} &= p^\mu u^\nu - u^\mu p^{\nu} \,,
\end{align}
where $p^\mu$ and $S^{\mu\nu}$ are the linear and angular momenta, respectively; $\tau$ is the proper time; $R^\mu{}_{\nu\rho\sigma}$ is the Riemann tensor; and $u^\nu \equiv \dv*{x^\nu}{\tau}$ is the four-velocity tangent to the worldline $x^\nu(\tau)$. To fix the worldline and thus to close the system of equations, we must choose an observer frame $V^\nu$ for which the worldline coincides with the center of mass, i.e., the mass dipole vanishes, $S^{\mu\nu}V_\nu = 0$. This is known as a spin supplementary condition (SSC). We use the Tulczyjew-Dixon \cite{tulczyjew1959motion,Dixon:1970I} SSC $V^\mu \propto p^\mu$.  

The spin magnitude $S=\sqrt{S^{\mu\nu} S_{\mu\nu}/2}$ and the mass $\mu = \sqrt{- p^\mu p_\mu}$ are conserved due to the choice of the SSC. We define the specific spin tensor as $s^{\mu\nu} = S^{\mu\nu}/\mu$. For an extremal Kerr black hole, the spin magnitude is $S = \mu^2$. For compact objects such as black holes or neutron stars, the specific spin magnitude $s=S/\mu$ satisfies $s \leq \mu = \epsilon M$, where $\epsilon$ is the mass ratio $\mu/M$. We neglect terms starting at the quadratic order in spin, which corresponds to truncating at the linear order in the small mass ratio. Then the four-momentum and four-velocity are parallel at the leading order, i.e. $p^\mu = \mu u^\mu$ \cite{Dixon:1970I}. After defining the specific spin vector $s^\mu \equiv -\epsilon^{\mu\nu\kappa\lambda} u_\nu s_{\kappa\lambda}/2$, $s^{\mu\nu} = \epsilon^{\mu\nu\kappa\lambda} u_\kappa s_\lambda$, 
where $\epsilon^{\mu\nu\kappa\lambda}$ is the Levi-Civita pseudotensor, we can write the equations of motion as 
\begin{align}
     \frac{D^2 x^\mu}{\dd \tau^2} &= - \frac{1}{2} R^{\mu}{}_{\nu\kappa\lambda} \epsilon^{\kappa\lambda}{}_{\rho\sigma} u^\nu u^\rho s^{\sigma} \,, & \frac{D s^{\mu}}{\dd \tau} &= 0 \,. \label{eq:MPDlinx}
\end{align}

%%%%%%%%%%%%%%%%%%%%%%%%%%%%%%%%%%%%%%%%%%%%%%%%%%%%%%%%%%%%%
\mytitle{Constants of motion}
We consider motion of spinning bodies in Kerr space-time where the line element in Boyer-Lindquist-like coordinates $(t, r, z = \cos\theta, \phi)$ reads as
\begin{multline}
    \dd s^2 = - \frac{\Delta}{\Sigma} \qty( \dd t - a (1-z^2) \dd \phi )^2 + \frac{\Sigma}{\Delta} \dd r^2 \\
    + \frac{\Sigma}{1-z^2} \dd z^2 + \frac{1-z^2}{\Sigma} \qty( a \dd t - (r^2 + a^2) \dd \phi )^2
\end{multline}
with $\Delta = r^2 - 2Mr + a^2$ and $\Sigma = r^2 + a^2 z^2$.

The Kerr space-time has a hidden symmetry represented by a conformal Killing-Yano tensor defined by the properties $f_{\mu \nu} = -f_{\nu\mu}$ and $f_{\mu\nu;\kappa} = (g_{\kappa \mu} f_{\lambda \nu}{}^{;\lambda}- g_{\kappa \nu} f_{\lambda \mu}{}^{;\lambda})/3$. Its components in Boyer-Lindquist coordinates read as
\begin{align}\label{eq:fmunu}
\begin{split}
    f_{\mu\nu} \dd x^\mu \wedge \dd x^\nu =
    & - r \, \dd r \wedge \qty( \dd t - a (1-z^2) \dd \phi ) 
    \\ & 
    + a z \, \dd z \wedge \qty( a \dd t - (r^2 + a^2) \dd \phi )\,.
\end{split}
\end{align}
From  various derivatives and contractions of this tensor one can derive the Killing vectors $\xi^\mu_{(t)}$ and $\xi^\mu_{(\phi)}$, the Killing-Yano tensor $Y_{\mu\nu} = \epsilon^{\kappa \lambda}{}_{\mu\nu} f_{\kappa\lambda}/2$, and the Killing tensor $K_{\mu\nu} = Y_{\mu\kappa} Y_\nu{}^\kappa$ \cite{Frolov:2017kze} (see also Table I in Supplemental Material \cite{supplemental}).

Due to the aforementioned symmetries, the following quantities are conserved to $\mathcal{O}(s)$ \cite{Dixon:1970I,Rudiger:1981,Rudiger:1983}
\begin{subequations}\label{eq:com_definitions}
\begin{align}
    E &= - \xi^{(t)}_\mu u^\mu + \frac{1}{2} \xi^{(t)}_{\mu;\nu} s^{\mu\nu} \, , \,\, J_z = \xi^{(\phi)}_\mu u^\mu - \frac{1}{2} \xi^{(\phi)}_{\mu;\nu} s^{\mu\nu} \, , \\
    K &= K_{\mu\nu} u^\mu u^\nu + 4 u^\mu s^{\rho\sigma} Y^\kappa{}_{\left[\mu\right.} Y_{\left.\sigma\right]\rho;\kappa}  \, , \label{eq:Carter} \\ 
    s_\parallel &= \frac{Y_{\mu\nu} u^\mu s^\nu}{\sqrt{K_{\mu\nu} u^\mu u^\nu}} \, .
\end{align}
\end{subequations}
These represent total specific energy; total specific azimuthal angular momentum; a Carter-like constant; and the projection of the specific spin vector onto the orbital angular momentum.

%%%%%%%%%%%%%%%%%%%%%%%%%%%%%%%%%%%%%%%%%%%%%%%%%%%%%%%%%%%%%
\mytitle{Geodesic order}
When the spin magnitude is zero, the MPD equations reduce to the equations of motion of a geodesic, which were shown to be separable by \citet{Carter:1968}. The components of the geodesic four-velocity $u^\mu_{\rm g}$ can be expressed as
\begin{subequations}\label{eq:geodesic}
\begin{align}
    \Sigma \dv{t}{\tau} &= T_r^{(E,J_z)}(r) + T_z^E(z) + a J_z \, , 
    \\
    \Sigma \dv{\phi}{\tau} &= \Phi_r^{(E,J_z)}(r) + \Phi_z^{J_z}(z) - a E \, ,
    \\
    \qty( \Sigma \dv{r}{\tau} )^2 &= R^{(E,J_z,K)}(r) \, , \;
    \qty( \Sigma \dv{z}{\tau} )^2 = Z^{(E,J_z,K)}(z) \, , 
\end{align}
\end{subequations}
where the functions on the right side can be found in \cite{Fujita:2009} and Supplemental Material \cite{supplemental}. After reparametrization with the Mino parameter $\lambda$ defined as $\dd \lambda = \Sigma^{-1} \dd \tau$ \cite{Mino:2003}, Eqs.~\eqref{eq:geodesic} separate into radial and polar parts and can be solved analytically. The analytical solution in terms of Legendre elliptic integrals and Jacobi and Weierstrass elliptic functions was found in \cite{Fujita:2009,vandeMeent:2020,Cieslik:2023}. 

For bound orbits, the motion is periodic in the Mino parameter with radial and polar frequencies $\Upsilon_{r,z}$ and average rates of change of $t$ and $\phi$ denoted by $\Upsilon_{t,\phi}$. The coordinate-time frequencies are given as $\Omega_{r,z,\phi}= \Upsilon_{r,z,\phi}/\Upsilon_t$.

\mytitle{Parallel transport}
In the linearized regime with nonzero spin, the spin vector affects the trajectory solely through the $\mathcal{O}(s)$ spin curvature term in Eq.~\eqref{eq:MPDlinx}. Thus, it can be evolved only at leading order accuracy, which corresponds to parallel transport of the vector along a geodesic. Using the solution for a parallel transported tetrad $e_A^\mu$ found by \citet{Marck:1983}, the spin vector is thus expressed as
\begin{equation}
    s^\mu = s_\parallel e_3^\mu + s_\perp \qty( e_1^\mu \cos(\psi) + e_2^\mu \sin(\psi) ) \, .
\end{equation} 
This form corresponds to a component $s_\parallel$ parallel to $e_3^\mu$ and another component $s_\perp \equiv \sqrt{s^2 - s_\parallel^2}$ precessing around $e_3^\mu$ with phase $\psi(\tau)$. The tetrad legs $e_A^\mu$ are constructed as follows. The zeroth leg is the four-velocity, but in the linear regime we can take the geodesic four-velocity $u^\mu_{\text{g}}$ defined in Eqs.~\eqref{eq:geodesic} with the constants of motion of spinning particle from Eqs.~\eqref{eq:com_definitions}. Therefore, the four-velocities $u^\mu$ and $u^\mu_{\text{g}}$ are $\order{s}$ close. The other components are obtained from $u^\mu_\text{g}$ and the Killing-Yano tensor $Y_{\mu\nu}$ as \cite{Witzany:2019}
\begin{subequations} \label{eq:marck}
\begin{align}
    e_1^\mu &= \frac{1}{N_{1}} \qty( K (Y^3)^{\mu}{}_\nu + K^{(2)} Y^\mu{}_\nu ) u^\nu_\text{g} \, , \\
    e_2^\mu &= - \frac{1}{N_{2}} \qty( (Y^2)^\mu{}_\nu - K \delta^\mu_\nu ) u^\nu_\text{g} \, , \\
    e_3^\mu &= -\frac{1}{\sqrt{K}} Y^\mu{}_\nu u^\nu_\text{g} \, ,
\end{align}
\end{subequations}
where
\begin{align}
    N_{1}^2 &= K^{(3)} K^2 - \qty(K^{(2)})^2 K \, , 
    \\
    N_{2}^2 &= K^{(2)} + K^2 \, , 
    \\
    K^{(n)} &= (-1)^n (Y^{2n})_{\mu\nu} u^\mu_\text{g} u^\nu_\text{g} \, , 
    \\
    (Y^n)^\mu{}_\nu &= \underbrace{Y^\mu{}_{\alpha} Y^\alpha{}_{\beta} \ldots Y^\gamma{}_\nu}_n \, . \label{eq:Ypower}
\end{align} 
The explicit form in Boyer-Lindquist coordinates is available in Supplemental Material \cite{supplemental}. 
With our choice of $e_1^\mu$ and $e_2^\mu$, the evolution equation for $\psi$ becomes separable \cite{Marck:1983}, $\mathrm{d} \psi/\mathrm{d} \lambda = -\Psi_r(r) - \Psi_z(z)$, where the functions $\Psi_r,\Psi_z$ are given in \cite{Piovano:2024} and the Supplemental Material. The phase $\psi$ as a function of Mino parameter in terms of Legendre elliptic integrals was given in \cite{vandeMeent:2020}, where $\psi$ is defined with an opposite sign. The corresponding Mino frequency is $\Upsilon_\psi$.

%%%%%%%%%%%%%%%%%%%%%%%%%%%%%%%%%%%%%%%%%%%%%%%%%%%%%%%%%%%%%
\mytitle{Worldline shift} 
We now transform to a worldline $\Tilde{x}^\mu(\tau)$ for which the equations of motion become separable. The worldline is defined as 
$ \Tilde{x}^\mu = x^\mu + \delta x^\mu $, 
where $\delta x^\mu = \order{s}$ is a displacement vector, which reads as
\begin{equation}
    \delta x^\mu = s_\parallel \delta x_3^\mu + s_\perp \qty( \delta x_1^\mu \cos(\psi) + \delta x_2^\mu \sin(\psi) ) \, ,
\end{equation}
where the vectors $\delta x^\alpha_a, a=1,2,3$ along with $u^\alpha_{\text{g}}$ form a basis of the tangent space and are given as
\begin{subequations}\label{eq:tetrad}
\begin{align}
    \delta x_1^\mu &= -\frac{1}{N_{1}} \qty( \mathcal{Q}^{(1)} (f^3)^{\mu}{}_\nu + \mathcal{Q}^{(2)} f^\mu{}_\nu ) u^\nu_\text{g} \, , \\
    \delta x_2^\mu &= \frac{a r z}{K N_{2}} \qty( (f^2)^\mu{}_\nu - \mathcal{Q}^{(1)} \delta^\mu_\nu ) u^\nu_\text{g} \, , \\
    \delta x_3^\mu &= -\frac{1}{\sqrt{K}} f^\mu{}_\nu u^\nu_\text{g} \,,
\end{align}
\end{subequations}
where  
\begin{align}
    \mathcal{Q}^{(n)} &= (-1)^n (f^{2n})_{\mu\nu} u^\mu_\text{g} u^\nu_\text{g} \, .
\end{align}
The Boyer-Lindquist coordinate components and some other properties of $\delta x^\mu_a$ are presented in Supplemental Material \cite{supplemental}. 

Vectors $\delta x^\mu_a$ are defined similarly to the Marck tetrad in Eqs.~\eqref{eq:marck}--\eqref{eq:Ypower}, using the conformal Killing-Yano tensor $f_{\mu\nu}$ instead of the Killing-Yano tensor $Y_{\mu\nu}$ (with an additional factor $a r z = f_{\mu\nu} Y^{\mu\nu}/4$ in $\delta x_2^\mu$). The normalization factors for the Marck tetrad are used in the definition of $\delta x^\mu_a$. Consequently, the vectors $u^\mu_\text{g},\delta x^\mu_{a}$ are mutually orthogonal, but the lengths of $\delta x^\mu_{a}$ are \textit{not} normalized to 1. 

\begin{figure}
    \centering
    \includegraphics{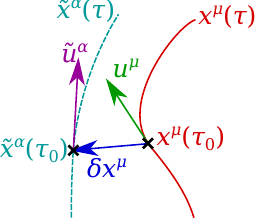}
    \caption{Worldline of the spinning particle $x^\mu(\tau)$ (bold red) with tangent vector $u^\mu$ (green) with the displacement vector $\delta x^\mu$ (blue) at point $x^\mu(\tau_0)$. This defines the new worldline $\Tilde{x}^\alpha(\tau)$ (dashed cyan) with the tangent vector $\Tilde{u}^\alpha$ (purple).
    }
    \label{fig:worldlines}
\end{figure}

The new equations of motion are derived by transforming the equations for the constants of motion \eqref{eq:com_definitions} to the new worldline. To do so, we employ Synge's worldfunctions and parallel propagators to transform the Killing vectors and tensor, the four-velocity, and the spin tensor (see Refs. \cite{Poisson:2011nh,Vines:2016unv}). Since we work in the linear regime, several aspects are simplified. For example, parallel propagators are $g^\mu{}_\alpha(x,\Tilde{x}) = \delta^\mu_{\alpha} + \order{s^2}$. Moreover, we do not have to transform tensors that are already multiplied by the spin tensor, since it would contribute only to quadratic and higher terms.

The Killing vectors and tensor transform as
\begin{align} 
    \xi^\mu &= g^\mu{}_{\alpha} \qty( \Tilde{\xi}^\alpha - \Tilde{\xi}^\alpha{}_{;\beta} \delta x^\beta + \order{s^2} )  \, ,\label{eq:transformation_xi} \\
    K_{\mu\nu} &= g_\mu{}^{\alpha} g_\nu{}^\beta \qty( \Tilde{K}_{\alpha\beta} - \Tilde{K}_{\alpha\beta;\gamma} \delta x^\gamma + \order{s^2} ) \, , \label{eq:transformation_K}
\end{align}
where $g^\mu{}_\alpha(x,\Tilde{x})$ is a parallel propagator\footnote{Here we use the indices $\mu$, $\nu, \ldots$ for the tensors in the tangent space at the original point and the indices $\alpha,\ldots,\delta$ for the tensors in the tangent space at the new point.} which parallel transports vectors from $\Tilde{x}^\alpha$ to $x^\mu$. $\Tilde{\xi}^\alpha$ and $\Tilde{K}_{\alpha\beta}$ denote the components of the Killing vectors and tensor in the new coordinates, respectively.

The new four-velocity $\Tilde{u}^{\alpha} = \dv*{\Tilde{x}^{\alpha}}{\tau}$ is the vector tangent  to the shifted worldline $\tilde{x}^\mu(\tau)$, as illustrated in Fig.~\ref{fig:worldlines}.  The original four-velocity $u^\mu = \dv*{x^\mu}{\tau}$ can be expressed as
\begin{equation}\label{eq:transformation_u}
    \dv{x^\mu}{\tau} = g^\mu{}_{\alpha} \qty( \dv{\Tilde{x}^{\alpha}}{\tau} - \frac{D \delta x^{\alpha}}{\dd \tau} + \order{s^2} )\,.
\end{equation}
 The original spin tensor $s^{\mu\nu}$ can be expressed using the spin tensor with respect to the new worldline $\Tilde{s}^{\alpha\beta}$ as
\begin{equation}\label{eq:transformation_S}
    s^{\mu\nu} = g^\mu{}_{\alpha} g^\nu{}_{\beta} \qty( \Tilde{s}^{\alpha\beta} - \Tilde{u}^{\alpha} \delta x^{\beta} + \Tilde{u}^{\beta} \delta x^{\alpha} + \order{s^2} ) \, .
\end{equation}
The spin tensor with respect to the new worldline acquires an additional mass dipole.  After transformation of $\Tilde{s}^{\alpha\beta}$ to a different frame while keeping the worldline $\Tilde{x}^\mu(\tau)$ fixed, the mass dipole generally does not vanish. As such, the worldline $\Tilde{x}^\mu(\tau)$ cannot be associated with any of the classical SSCs $\Tilde{s}^{\mu\nu} V_\nu = 0$ because the spin tensor $\Tilde{s}^{\mu\nu}$ is not degenerate. This is illustrated for fully parallel spin in Supplemental Material.

\begin{figure}
    \centering
    \includegraphics[height=4cm]{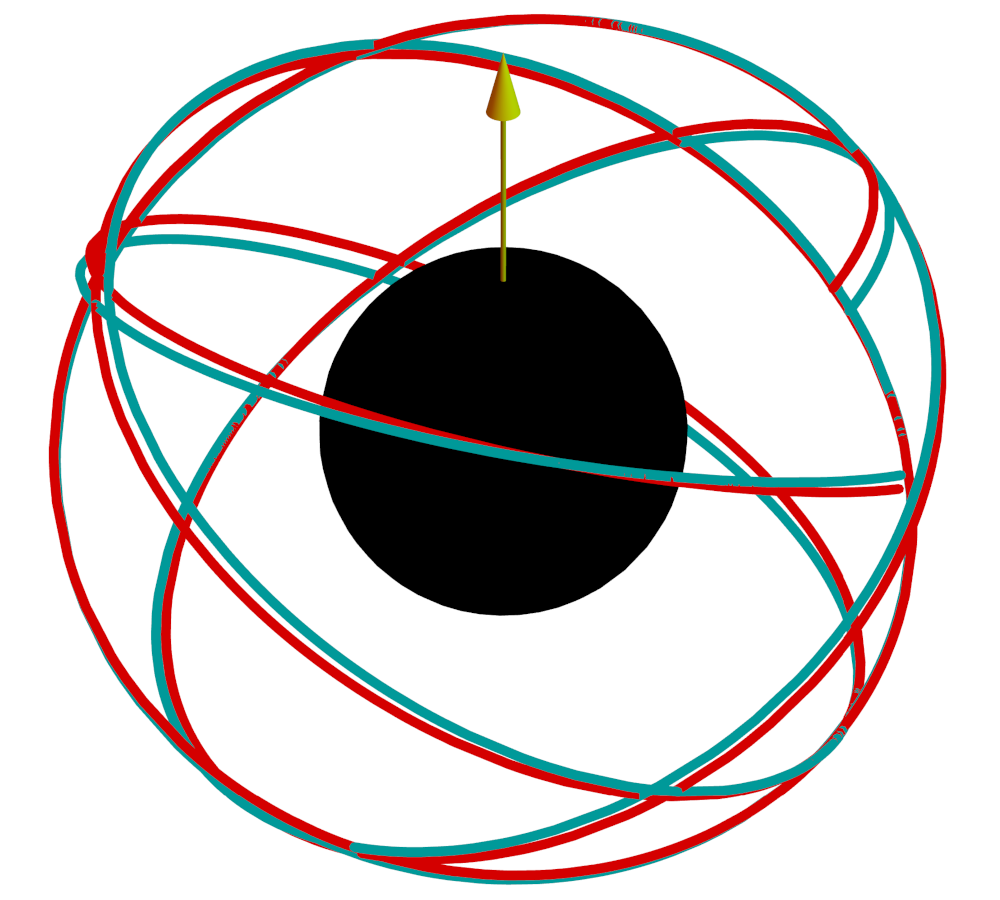}
    \includegraphics[height=4cm]{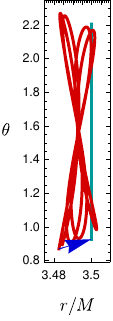}
    \caption{Trajectory of a spinning particle $x^\mu(\tau)$ (red) and the virtual worldline $\Tilde{x}^\mu(\tau)$ (cyan) in a 3D plot (left) and projection into the $r-\theta$ plane (right). Primary spin is indicated with a yellow arrow while the shift vector at $\tau = 0$ is indicated by a blue arrow. The orbital parameters correspond to a nearly spherical orbit (i.e., the virtual worldline is spherical). Spin components $s_\parallel=0.005M,\, s_\perp = 0.1M$ are exaggerated to make the two trajectories distinguishable.}
    \label{fig:enter-label}
\end{figure}

%---------------------------------------------------------------------------------------
\mytitle{Separation of the equations of motion} 
After applying the transformations \eqref{eq:transformation_xi} -- \eqref{eq:transformation_S}, the constants of motion \eqref{eq:com_definitions} and the normalization of four-velocity can be expressed as
\begin{subequations}\label{eq:constants2}
\begin{align}
    E &= - \Tilde{\xi}_{(t)}^{\alpha} \qty( \Tilde{u}_{\alpha} - \frac{D \delta x_{\alpha}}{\dd \tau} ) + \frac{1}{2} \Tilde{\xi}^{(t)}_{\alpha;\beta} \Tilde{s}^{\alpha\beta} \, , \\
    J_z &= \Tilde{\xi}_{(\phi)}^{\alpha} \qty( \Tilde{u}_{\alpha} - \frac{D \delta x_{\alpha}}{\dd \tau} ) - \frac{1}{2} \Tilde{\xi}^{(\phi)}_{\alpha;\beta} \Tilde{s}^{\alpha\beta} \, , \\
    K &= \Tilde{K}_{\alpha\beta} \Tilde{u}^{\alpha} \qty( \Tilde{u}^{\beta} - 2 \frac{D \delta x^{\beta}}{\dd \tau} ) + 4 \Tilde{u}^{\alpha} \Tilde{s}^{\beta\gamma} \Tilde{Y}^{\delta}{}_{[\alpha} \Tilde{Y}_{\gamma]\beta;\delta} \, , \\
    -1 &= \Tilde{g}_{\alpha\beta} \Tilde{u}^{\alpha} \qty( \Tilde{u}^{\beta} - 2 \frac{D \delta x^{\beta}}{\dd \tau} ) \, .
\end{align}
\end{subequations}
We define an auxilliary vector,
\begin{equation}\label{eq:vec_v}
    \Tilde{v}^{\alpha} = \Tilde{u}^{\alpha} \qty(1 - \frac{3 s_\parallel E}{2 \sqrt{K}}) + \frac{3 s_\parallel}{2\sqrt{K}} \Tilde{\xi}_{(t)}^{\alpha}\,,
\end{equation}  
to substitute for $\Tilde{u}^{\alpha}$ in Eqs.~\eqref{eq:constants2}. After calculating the spin terms in Eqs.~\eqref{eq:constants2}, the perpendicular components of spin containing the precession phase $\psi$ cancel out. This behavior is consistent with the independence of \textit{averaged} observables on the precession phase to the linear order \cite{Witzany:2019,Drummond:2022a,Drummond:2022b,Piovano:2024}. However, this is the first time where such a fact was observed even in a \textit{local-in-time} expression of the motion.

The equations can then be reorganized into the relations\footnote{See the Supplemental Material \cite{supplemental} for a \textit{Mathematica} notebook containing the calculations.}
\begin{subequations}\label{eq:new_constants}
\begin{align}
    - \Tilde{\xi}_{(t)}^\alpha \Tilde{v}_\alpha &= E + \frac{s_\parallel(1 - E^2)}{2\sqrt{K}} \equiv \Tilde{E} \, ,  \\
    \Tilde{\xi}_{(\phi)}^\alpha \Tilde{v}_\alpha &= J_z + \frac{s_\parallel(a - J_z E/2)}{\sqrt{K}} \equiv \Tilde{J}_z \, ,  \\
    \Tilde{K}_{\alpha\beta} \Tilde{v}^\alpha \Tilde{v}^\beta &= K + \frac{s_\parallel( 3 a (J_z-aE) - K E )}{\sqrt{K}} \equiv \Tilde{K} \, , \\
    \Tilde{g}_{\alpha\beta} \Tilde{v}^\alpha \Tilde{v}^\beta &= - 1 \, .
\end{align}
\end{subequations}
This system of equations is formally identical to the relations for a geodesic with the constants of motion $(\Tilde{E}, \Tilde{J}_z, \Tilde{K})$ and four-velocity $\Tilde{v}^\alpha$. As such, $\Tilde{v}^\alpha$ can be expressed using Eqs.~\eqref{eq:geodesic} with $(\Tilde{E}, \Tilde{J}_z, \Tilde{K})$ instead of $({E}, {J}_z, {K})$. Finally, we can express the components of the new four-velocity $\Tilde{u}^\alpha$ from Eq.~\eqref{eq:vec_v} using the components of $\Tilde{v}^\alpha$ as
\begin{subequations}\label{eq:eom_final}
\begin{align}
    \dv{\Tt}{\Tl} &= T_r^{(\Tilde{E},\Tilde{J}_z)}(\Tr) + T_z^{\Tilde{E}}(\Tz) + a \Tilde{J}_z - \frac{3 s_\parallel \Tilde{\Sigma}}{2\sqrt{K}} \, , \label{eq:eom_t}\\
    \qty( \dv{\Tr}{\Tl} )^2 &= R^{(\Tilde{E}, \Tilde{J}_z, \Tilde{K})}(\Tr) \, , \\
    \qty( \dv{\Tz}{\Tl} )^2 &= Z^{(\Tilde{E}, \Tilde{J}_z, \Tilde{K})}(\Tz) \, , \\
    \dv{\Tphi}{\Tl} &= \Phi_r^{(\Tilde{E}, \Tilde{J}_z)}(\Tr) + \Phi_z^{\Tilde{J}_z}(\Tz) - a \Tilde{E} \, ,
\end{align}    
\end{subequations}
where we introduced a deformed Mino parameter,
\begin{equation}\label{eq:eom_tau_final}
    \dv{\tau}{\Tl} = \qty(1 - \frac{3 s_\parallel E}{2 \sqrt{K}}) \Tilde{\Sigma}\,,
\end{equation}
and where $\Tilde{\Sigma} = \Sigma(\Tr,\Tz)$. Identical equations can be derived directly by performing a change of variables in the equations of motion for the original worldline $x^\mu$ (see Supplemental Material).

The term in Eq.~\eqref{eq:eom_t} originating from the Killing vector $\Tilde{\xi}^\alpha_{(t)}$ that contains the function $\Tilde{\Sigma}$ is proportional to the equation for the geodesic $\dv*{\tau}{\lambda}$, and the solution can be written using the geodesic solution for proper time. Thus, the final trajectory can be expressed using the geodesic trajectory with a shift in the constants of motion, displacement vector $\delta x^\mu = (\delta t, \delta r, \delta z, \delta \phi)$, and geodesic proper time from the Killing vector term, as
\begin{subequations}\label{eq:result}
\begin{align}
    t(\Tl; C) &= t_\text{g}(\Tl; \Tilde{C}) - \frac{3 s_\parallel}{2 \sqrt{K}} \tau_\text{g}(\Tl; \Tilde{C}) - \delta t(\Tl) \, , \\
    x^k(\Tl; C) &= x^k_\text{g}(\Tl; \Tilde{C}) - \delta x^k(\Tl) \, , \\
    \tau(\Tl; C) &= \qty( 1 - \frac{3 s_\parallel E}{2 \sqrt{K}} ) \tau_\text{g}(\Tl; \Tilde{C}) \, , \label{eq:result_tau}
\end{align}
\end{subequations}
where $\delta x^\mu(\Tl) \equiv \delta x^\mu(r_\text{g}(\Tl),z_\text{g}(\Tl),\psi(\Tl))$, the ``g'' subscript denotes geodesic quantities which can be found in \cite{Fujita:2009,vandeMeent:2020,Cieslik:2023}, $C = (E,J_z,K)$, and $\Tilde{C} = (\Tilde{E},\Tilde{J}_z,\Tilde{K})$. Here, $t_\text{g}(\_ \ ;\_)$, $x^k_\text{g}(\_ \ ;\_)$, and $\tau_\text{g}(\_ \ ;\_)$ do not refer to any particular worldline, but must be understood purely as functional expressions.

Similarly, the Mino frequencies of the transformed coordinates can be expressed using the geodesic quantities as
\begin{subequations}\label{eq:Mino_frequencies_result}
\begin{align}
    \Upsilon_t(C) &= \Upsilon_t^\text{g}(\Tilde{C}) - \frac{3 s_\parallel}{2\sqrt{K}} \Upsilon_\tau^\text{g}(\Tilde{C}) \, , \\
    \Upsilon_k(C) &= \Upsilon_k^\text{g}(\Tilde{C}) \, , \\
    \Upsilon_\tau(C) &= \qty( 1 - \frac{3 s_\parallel E}{2 \sqrt{K}} ) \Upsilon_\tau^\text{g}(\Tilde{C}) \, .
\end{align}    
\end{subequations}
This holds even for the time and azimuthal frequencies because the shifts $\delta t$ and $\delta \phi$ are purely oscillatory. However, radial and polar motion is not periodic in $\Tl$ since the frequency spectrum contains the other frequencies from the shift $\delta r$ and $\delta z$.  Coordinate-time or proper-time frequencies can be calculated analogously to the geodesic case by dividing the Mino frequencies by $\Upsilon_t$ or $\Upsilon_\tau$.

%%%%%%%%%%%%%%%%%%%%%%%%%%%%%%%%%%%%%%%%%%%%%%%%%%%%%%%%%%%%%
\mytitle{Properties of the solution}
The $\delta x^\mu_{1,2}$ components of the shift vector can be viewed as ``integrating out'' all the oscillations of the trajectory caused by the $\psi$-dependent oscillations of the $\propto s_\perp$ component of spin. In fact, these components project the trajectory into a plane orthogonal to the total angular-momentum vector in Schwarzschild space-time (see \cite{Witzany:2023}). The component $\delta x^\mu_3$ then deals with the persistent nonseparability due to the $s_\parallel$ component.

Unlike some of the other formalisms, there are no particular issues when approaching near-equatorial or near-spherical orbits.
Since they are related by a bounded shift vector, the virtual orbit must lose stability (begin to plunge) at the same time as the MPD trajectory. We can thus find the $\mathcal{O}(s)$ correction to the separatrix by using the relations for geodesics \cite{Stein:2019buj} and plugging in the tilde constants $\Tilde{C}$.

%%%%%%%%%%%%%%%%%%%%%%%%%%%%%%%%%%%%%%%%%%%%%%%%%%%%%%%%%%%%%
\mytitle{Relation to other results} 
The Mino frequencies \eqref{eq:Mino_frequencies_result} are different from the frequencies calculated by \citet{Drummond:2022b} and \citet{Piovano:2024}, which we denote as $\Upsilon_\mu^{\text{DH}}$. The reason is that while in \cite{Drummond:2022b,Piovano:2024} the relation for Mino parameter reads $\dd \tau = \Sigma \dd \lambda$, here we use the relation \eqref{eq:eom_tau_final} after the transformation to $\Tilde{x}^\mu$ and with the factor $(1-3s_\parallel E/(2\sqrt{K}))$. Therefore, the relation between our parameter $\Tl$ and Mino parameter in \cite{Drummond:2022b,Piovano:2024} reads as
\begin{equation}
    \dv{\lambda}{\Tilde{\lambda}} 
    = \dv{\tau}{\Tl} \qty( \dv{\tau}{\lambda} )^{-1} 
    = \qty( 1 - \frac{3 s_\parallel E}{2\sqrt{K}} ) \frac{\Tr^2 + a^2 \Tz^2}{r^2 + a^2 z^2} \, .
\end{equation}
The explicit expression can be found in Supplemental Material. The frequencies then transform as
\begin{equation}
    \Upsilon_\mu = \left\langle \dv{\lambda}{\Tl} \right\rangle \Upsilon_\mu^{\text{DH}} \, ,
\end{equation}
where the angle brackets denote averaging. After this transformation, the frequencies calculated in \cite{Drummond:2022b} agree with our approach up to the linear order in spin. Finally, the frequencies $\Upsilon_\mu$ are related to the frequencies derived using the solution of the Hamilton-Jacobi equation in \cite{Witzany:2024ttz} $\Upsilon_\mu^{\text{HJ}}$ as $\Upsilon_\mu = (1 - s_\parallel E/(2\sqrt{K})) \Upsilon_\mu^{\text{HJ}}$.

%%%%%%%%%%%%%%%%%%%%%%%%%%%%%%%%%%%%%%%%%%%%%%%%%%%%%%%%%%%%%
\mytitle{Comparison with numerically calculated trajectories}
We verified the analytical formulas for the trajectories by comparing their result with trajectories calculated by numerically integrating the MPD equations \eqref{eq:MPD_full}. Details of the matching are given in Supplemental Material. The comparison revealed that our analytical result differs from the numerical one by $\order{s^2}$, which is consistent with the spin linearization.

%%%%%%%%%%%%%%%%%%%%%%%%%%%%%%%%%%%%%%%%%%%%%%%%%%%%%%%%%%%%%
\mytitle{Extensions of this work} This work could be extended by examining the expressions \eqref{eq:new_constants} in tensor notation without evaluating the tensor components in coordinates.
This may reveal the connection between the (conformal) Killing-Yano tensor and the shift transform. This would then facilitate the expression of the result for
other space-times with hidden symmetry such as the Kerr-NUT-(A)dS family in four and higher dimensions \cite{Carter:1968ks,Frolov:2017kze}. 

Another potential generalization would be to separate the equations of motion of pole-dipole-quadrupole particles by applying a second-order shift in $s$ \cite{Vines:2016unv}. There are enough constants of motion for the integrability of the pole-dipole-quadrupole system if the spin-induced quadrupole of the ``particle'' corresponds to the Geroch-Hansen quadrupole of a Kerr black hole \cite{Compere:2023alp,Ramond:2024ozy}.\footnote{The value of the quadrupole also corresponds to $\mathcal{N}=2$ supersymmetry along the wordline \cite{Jakobsen:2021lvp}.} This suggests that separability is possible, though finding the appropriate shift vector may be challenging.

%%%%%%%%%%%%%%%%%%%%%%%%%%%%%%%%%%%%%%%%%%%%%%%%%%%%%%%%%%%%%
\mytitle{Uses of the analytical solution}
The most straightforward use of our work will be to couple the analytical solution to a Teukolsky solver in order to generate fluxes of gravitational waves sourced by the motion, similarly to Refs \cite{Skoupy:2023lih,Piovano:2024}. The analytical solution will evaluate faster than semi-analytical ones,\footnote{This holds because the precalculation of the Fourier coefficients for the semi-analytic and numerical solution \cite{Piovano:2024,Drummond:2022b} takes seconds to tens of seconds. Additionally, the Jacobi elliptic functions are faster to evaluate than the Fourier series by a factor of 2 to 5.
} and there may be additional simplifications in the Teukolsky computation due to hidden symmetry. Additionally, the solution can be used to compute the evolution of the Carter-like constant $K$ under radiation reaction using methods akin to \citet{Sago:2005gd}. Finally, the test particle solution will provide a limiting reference for spinning binary dynamics at finite mass ratios \cite{Barausse:2009xi,Damour:2014sva} and a classical limit for scattering of quantum fields with nonzero spin \cite{Bern:2020buy,Jakobsen:2023ndj}.

%%%%%%%%%%%%%%%%%%%%%%%%%%%%%%%%%%%%%%%%%%%%%%%%%%%%%%%%%%%%%
\mytitle{Open questions} The separability of the Hamilton-Jacobi equation for geodesic motion follows from the separability of the scalar wave equation in Kerr space-time by taking an eikonal Wentzel–Kramers–Brillouin limit. Conversely, the Hamilton-Jacobi equation for the motion of spinning particles in Kerr is ``almost'' separable in Boyer-Lindquist coordinates when a Marck-tetrad basis is used for spin components \cite{Witzany:2019}. Independently, the massive spinor wave equation in Kerr space-time is separable in Kerr space-time \cite{Chandrasekhar:1976ap}. What are the interrelations between the previous (almost-)separability results for spinning particles and the current work? Can a canonical transformation counterpart to our worldline shift yield a fully separable Hamilton-Jacobi equation for spinning particles? Is there an eikonal limit of the spinor equation (such as that in Refs. \cite{audretsch1981trajectories, Rudiger:1981uu}) that would provide the corresponding solution of the spinning-particle Hamilton-Jacobi equation? Are such integrability and separability results limited to test fields and particles in Kerr space-time, or do they also apply when considering backreaction on the background? These questions warrant future investigation.

%%%%%%%%%%%%%%%%%%%%%%%%%%%%%%%%%%%%%%%%%%%%%%%%%%%%%%%%%%%%%
\mytitle{Acknowledgements}
We thank the Charles U. \textit{Primus} Research Program 23/SCI/017 for support. This work makes use of the Black Hole Perturbation Toolkit \cite{BHPToolkit}, specifically the GeneralRelativityTensors and KerrGeodesics packages \cite{KerrGeodesics,GeneralRelativityTensors}. VS thanks Georgios Lukes-Gerakopoulos for providing the code for the numerical calculation of MPD equations. We also thank Lisa Drummond and Gabriel Piovano for reading through a draft of this paper and giving us helpful feedback.

\bibliography{main}

\newpage
 
\include{SupplMat}

\end{document}

%% file: SupplMat.tex
\appendix
\onecolumngrid
\renewenvironment{widetext}{}{}

\setcounter{equation}{0}
\setcounter{figure}{0}
\renewcommand{\theequation}{S\arabic{equation}}
\renewcommand{\thefigure}{S\arabic{figure}}

\section{\normalsize Supplemental material to Skoup{\'y} \& Witzany (2024) \em Analytic solution for the motion of spinning particles in Kerr space-time }

\subsection{Functions in geodesic equations of motion}
The functions appearing in the separated geodesic equations \eqref{eq:geodesic} and in the evolution of the precession phase $\psi$ in the main text are
\begin{align}
    T_r^{(E,J_z)}(r) & \equiv \frac{r^2 + a^2}{\Delta} P_r(r) \,, & T_z^E(z) &\equiv -a^2 E (1 - z^2)\,, \label{eq:geodesic_Tr}\\
    R^{(E,J_z,K)}(r) & \equiv P_r(r)^2 - \Delta (K + r^2)\,,& Z^{(E,J_z,K)}(z) &\equiv (1 - z^2)(K -a^2 z^2) - P_z(z)^2 \,,\\
    \Phi_r^{(E,J_z)}(r) & \equiv \frac{a}{\Delta} P_r(r)\,, & \Phi_z^{J_z}(z) &\equiv \frac{J_z}{1 - z^2}\,, \label{eq:geodesic_Phiz}\\
    \Psi_r(r) &\equiv \frac{\sqrt{K} P_r(r)}{K + r^2} \,, & \Psi_z(z) &\equiv \frac{a \sqrt{K} P_z(z)}{K - a^2 z^2} \,,
\end{align}
where
\begin{align}
    \Delta &= r^2 - 2 M r + a^2 \, , & \Sigma &= r^2 + a^2 z^2 \, , \\
    P_r &= (r^2+a^2)E - a J_z \, , & P_z &= J_z - a E (1-z^2) \, .
\end{align}
\citet{Fujita:2009} use a modified Carter constant $\mathcal{C}$, while we use the original Carter constant $K = \mathcal{C} + (J_z - a E)^2$. The function $\Theta(\cos \theta)$ in \cite{Fujita:2009} is our $Z^{(E,J_z,K)}(z)$ when substituting $z=\cos \theta$.
\subsection{Coordinate expression of the displacement vector}
The components of the Marck tetrad we use in this work read
\begin{align}
    - e^0_\mu \dd x^\mu &= - E \dd t + \frac{1}{\Delta} \dv{r}{\lambda} \dd r + \frac{1}{1-z^2} \dv{z}{\lambda} \dd z + J_z \dd \phi \, , \\
    e^1_\mu \dd x^\mu &= \frac{-r \Xi \dv{r}{\lambda} + a^2 z \Xi^{-1} \dv{z}{\lambda}}{\sqrt{K} \Sigma} \dd t + \frac{r \Xi P_r}{\sqrt{K} \Delta} \dd r + \frac{a z \Xi^{-1} P_z}{\sqrt{K} (1-z^2)}\dd z + a \frac{r(1-z^2)\Xi \dv{r}{\lambda} - z(r^2+a^2)\Xi^{-1} \dv{z}{\lambda}}{\sqrt{K} \Sigma} \dd \phi \, , \\
    e^2_\mu \dd x^\mu &= -\frac{\Xi P_r + a \Xi^{-1} P_z}{\Sigma} \dd t + \frac{\Xi}{\Delta} \dv{r}{\lambda} \dd r + \frac{\Xi^{-1}}{1-z^2} \dv{z}{\lambda} \dd z + \frac{a(1-z^2) \Xi P_r + (a^2+r^2) \Xi^{-1} P_z}{\Sigma} \dd \phi \, , \\
    e^3_\mu \dd x^\mu &= a\frac{z \dv{r}{\lambda} + r \dv{z}{\lambda}}{\sqrt{K} \Sigma} \dd t + \frac{a z P_r}{\sqrt{K} \Delta} \dd r + \frac{r P_z}{\sqrt{K} (1-z^2)} \dd z - \frac{a^2 z (1-z^2) \dv{r}{\lambda} + r (r^2+a^2) \dv{z}{\lambda}}{\sqrt{K} \Sigma} \dd \phi \, ,
\end{align}
where
\begin{align}
    \dv{r}{\lambda} = \pm \sqrt{R^{(E,J_z,K)}(r)} \, , \quad \dv{z}{\lambda} = \pm \sqrt{Z^{(E,J_z,K)}(z)} \,, \quad
    \Xi = \sqrt{\frac{K - a^2 z^2}{K + r^2}} \, . 
\end{align}
This tetrad is equivalent to the tetrad of \citet{Witzany:2019} up to the signs of $e^1_\mu$, $e^2_\mu$, and $e^3_\mu$ and the order of $e^1_\mu$ and $e^2_\mu$, and to the tetrad of \citet{vandeMeent:2020} up to the sign of $e^2_\mu$ and $e^3_\mu$.

Similarly, spacelike components of the tetrad $(u^\mu_\text{g}, \delta x_1^\mu, \delta x_2^\mu, \delta x_3^\mu)$ can be expressed as
\begin{align}
    \delta x_1^\mu \partial_\mu &= \frac{1}{\sqrt{K} \Sigma} \qty( a \qty(\frac{a^2+r^2}{\Delta} z \Xi \dv{r}{\lambda} + r \Xi^{-1} \dv{z}{\lambda} ) \partial_t + a z \Xi P_r \partial_r  - r \Xi^{-1} P_z \partial_z + \qty( \frac{a^2 z}{\Delta} \Xi \dv{r}{\lambda} + \frac{r}{(1-z^2)} \Xi^{-1} \dv{z}{\lambda} ) \partial_\phi ) \, , \\
    \delta x_2^\mu \partial_\mu &= - \frac{a r z}{K \Sigma} \qty( \qty( \frac{r^2+a^2}{\Delta} \Xi P_r + a \Xi^{-1} P_z ) \partial_t + \Xi \dv{r}{\lambda} \partial_r + \Xi^{-1} \dv{z}{\lambda} \partial_z + \qty( \frac{a}{\Delta} \Xi P_r + \frac{1}{1-z^2} \Xi^{-1} P_z ) \partial_\phi ) \, , \\
    \delta x_3^\mu \partial_\mu &= \frac{1}{\sqrt{K} \Sigma} \qty( \qty( r \frac{r^2+a^2}{\Delta} \dv{r}{\lambda} - a^2 z \dv{z}{\lambda} ) \partial_t + r P_r \partial_r + a z P_z \partial_z + a \qty( \frac{r}{\Delta} \dv{r}{\lambda} - \frac{z}{1-z^2} \dv{z}{\lambda} ) \partial_\phi ) \, .
\end{align}
Note that $s_\perp \delta x_1^z \sin(\psi)$ in the Schwarzschild space-time corresponds to Eq.~(26) in \cite{Witzany:2023}. Furthermore, $s_\perp \delta x_1^z \sin(\psi)$ also corresponds to Eq.~(3.81) in \cite{Piovano:2024} in the near-equatorial limit in Kerr. The normalizations of the tetrad read
\begin{equation}
    \delta x_1^\mu \delta x^1_\mu = \delta x_3^\mu \delta x^3_\mu = 1 + \frac{r^2 - a^2 z^2}{K} \, , \qquad \delta x_2^\mu \delta x^2_\mu = \frac{a^2 r^2 z^2}{K^2} \, ,
\end{equation}
while the projections to the Marck tetrad read
\begin{equation}
    e^A_\mu \delta x^\mu_B = \mqty( 1 & 0 & 0 & 0 \\
                                    0 & - \frac{a r z}{K} & 0 & \frac{\sqrt{(K + r^2)(K-a^2 z^2) }}{K} \\
                                    0 & 0 & - \frac{a r z}{K} & 0 \\
                                    0 & -\frac{\sqrt{(K + r^2)(K-a^2 z^2) }}{K} & 0 & - \frac{a r z}{K} ) \, ,
\end{equation}      
where $A$ is the row index and $B$ is the column index.

\begin{table}[t]
    \centering
    \bgroup
    \def\arraystretch{1.5}
    \begin{tabular}{c|c}
        Definition & Property \\ \hline
        $\displaystyle f_{\mu\nu}$ (Eq.~\eqref{eq:fmunu}) & $\displaystyle f_{\mu\nu;\kappa} = g_{\kappa\mu} \xi^{(t)}_\nu - g_{\kappa\nu} \xi^{(t)}_{\mu}$ \\
        $\displaystyle \xi_{(t)}^\mu = \frac{1}{3} f^{\nu\mu}{}_{;\nu} = \delta^\mu_t $ & $\displaystyle \xi^{(t)}_{(\mu;\nu)} = 0$ \\
        $\displaystyle Y_{\mu\nu} = \frac{1}{2} f_{\kappa\lambda} \epsilon^{\kappa\lambda}{}_{\mu\nu}$ & $\displaystyle Y_{\mu(\nu;\rho)} = 0$ \\
        $\displaystyle K_{\mu\nu} = Y_{\mu\kappa} Y_{\nu}{}^\kappa$ & 
        $\displaystyle K_{(\mu\nu;\rho)}  = 0$
        \\
        $\displaystyle \zeta^\mu = K^{\mu\nu} \xi^{(t)}_\nu$ & $\displaystyle \zeta_{(\mu;\nu)} = 0$ \\
        $\displaystyle \xi^\mu_{(\phi)} = a^{-1} \zeta^\mu - a \xi_{(t)}^\mu = \delta^\mu_\phi$ & $\displaystyle \xi_{(\mu;\nu)}^{(\phi)} = 0$
    \end{tabular}
    \egroup
    \caption{Definition and properties of the conformal Killing-Yano tensor $f_{\mu\nu}$, primary and secondary Killing  vectors $\xi_{(t)}^\mu$ and $\zeta^\mu$, Killing-Yano tensor $Y_{\mu\nu}$, Killing tensor $K_{\mu\nu}$, and azimuthal Killing vector $\xi^\mu_{(\phi)}$.}
    \label{tab:Killing_tensors}
\end{table}

From the identity for the conformal Killing-Yano tensor in Table \ref{tab:Killing_tensors} we can simplify the expressions \eqref{eq:transformation_u} and \eqref{eq:transformation_S} for the parallel part. In particular, it holds
\begin{align}
    \frac{D \delta x_3^{\alpha}}{\dd \tau} &= -\frac{1}{\sqrt{K}} f^{\alpha}{}_{\beta;\gamma} u^\beta_\text{g} \Tilde{u}^\gamma = -\frac{1}{\sqrt{K}} \qty( \xi_{(t)}^\alpha - E u^\alpha_\text{g} + \order{s} ) \, , \\
    \Tilde{s}^{\alpha\beta}_\parallel &= s_\parallel \qty(\Tilde{\epsilon}^{\alpha\beta\gamma\delta} u_\gamma e^3_\delta + \Tilde{u}^\alpha \delta x_3^\beta - \Tilde{u}^\beta \delta x_3^\alpha ) = - \frac{s_\parallel}{\sqrt{K}} f^{\alpha\beta} \, .
\end{align}
Using this simple expression we see that the worldline does not corresponds to any SSC of the form $\tilde{s}^{\alpha \beta} V_\beta = 0$ with $V_\beta$ some time-like frame vector. This is because the tensor $f^{\alpha \beta}$ is non-degenerate for non-zero BH spin and thus there exists no $V_\beta$ that could generally appear in the SSC. As such, the worldline $\tilde{x}^\mu$ will be interpreted as shifted away from the center of mass of the spinning body in every observer frame.

The explicit expression for the transformation of the Mino parameter reads
\begin{equation}\label{eq:deltalambda}
    \dv{\lambda}{\Tl} = 1 + \frac{s_\parallel}{\sqrt{K}} \qty( -\frac{3}{2} E + 2 \frac{r^2 P_r + a^3 z^2 P_z }{\Sigma^2} ) + \frac{2 s_\perp a r z}{\Sigma^2 K} \qty( \sqrt{K} \qty( \Xi P_r - a \Xi^{-1} P_z ) \cos(\psi) - \qty( r \Xi \dv{r}{\lambda} + a^2 z \Xi^{-1} \dv{z}{\lambda} ) \sin(\psi) ) \, .
\end{equation}
After averaging this expression over the radial and polar motion, the perpendicular part vanishes. The term proportional to $\Sigma^{-2}$ cannot be separated and expressed analytically using standard methods.

The expression \eqref{eq:deltalambda} was used only for comparison with other results. For practical applications of our analytical solution, such as the calculation of gravitational-wave fluxes, it is possible to work with the $\Tl$ parametrization without the transformation to $\lambda$.

\begin{figure}[t]
    \centering
    \includegraphics[height=6.cm]{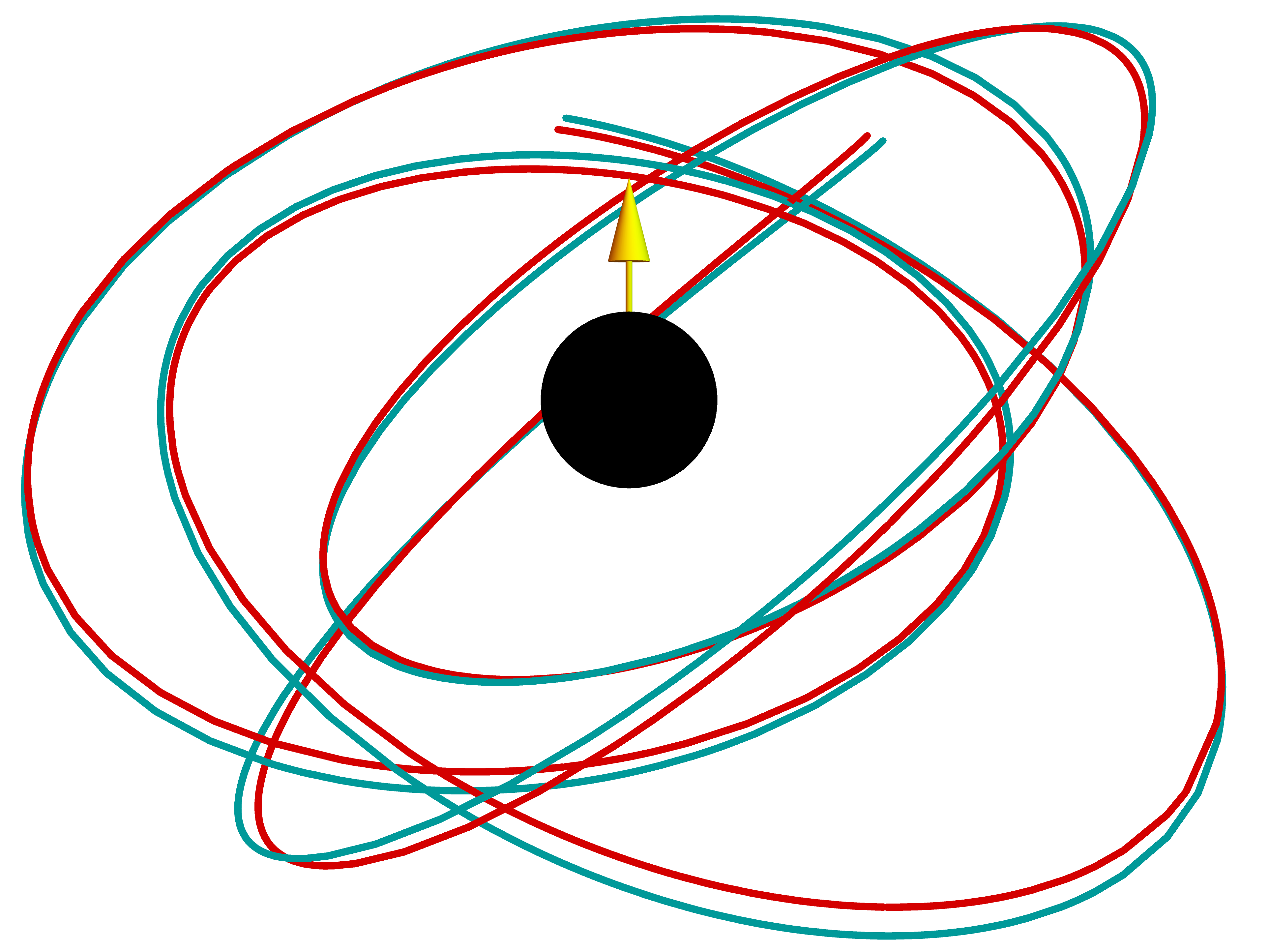}
    \includegraphics[height=6.cm]{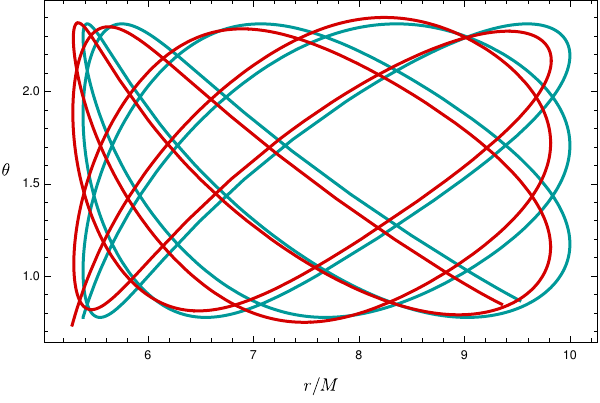}
    \caption{Trajectory of spinning particle $x^\mu(\tau)$ (red) and the virtual worldline $\Tilde{x}^\mu(\tau)$ (cyan) in a 3D plot (left) and projection into $r-\theta$ plane (right). The constants of motion of are $E = 0.9369$, $J_z = 2.1637 M$, $K = 6.6491 M^2$, $s_\parallel = 0.05 M$ and $s_\perp = 0.1M$. The magnitude of the secondary spin is unphysically large to make the trajectories easily distinguishable. We can see on the right plot that, while the worldline $\Tilde{x}^\mu$ reminds a geodesic trajectory, due to the shift vector $\delta x^\mu$, the real worldline $x^\mu$ has modulated turning points and the radius is shifted on average. }
    \label{fig:trajectory_generic}
\end{figure}

\subsection{Direct derivation in coordinates}

The separated equations of motion can be directly derived by performing a change of variables in the non-separable equations of motion. We start by inverting Eqs.~\eqref{eq:com_definitions} to obtain the components of the four-velocity
\begin{subequations}\label{eq:eom_nonseparable}
\begin{align}
    \Sigma \dv{t}{\tau} &= T_r^{(E,J_z)}(r) + T_z^E(z) + a J_z + \delta T(r,z,\psi) + \order{s^2} \,, \\
    \Sigma \dv{\phi}{\tau} &= \Phi_r^{(E,J_z)}(r) + \delta \Phi(r,z,\psi) + \order{s^2} \,, \\
    \qty( \Sigma \dv{r}{\tau} )^2 &= R^{(E,J_z,K)}(r) + \delta R(r,z,\psi) + \order{s^2} \,, \\
    \qty( \Sigma \dv{z}{\tau} )^2 &= Z^{(E,J_z,K)}(z) + \delta Z(r,z,\psi) + \order{s^2} \,,
\end{align}
\end{subequations}
where we used the geodesic functions from Eqs.~\eqref{eq:geodesic_Tr}--\eqref{eq:geodesic_Phiz}. The linear-in-spin corrections can be split as $\delta X = s_\parallel \delta X_\parallel + s_\perp \cos(\psi) \delta X_{\rm c} + s_\perp \sin(\psi) \delta X_{\rm s}$, where $X=T,\Phi,R,Z$. After some algebraic simplifications, these functions can be expressed as
\begin{subequations}\label{eq:delta_eom}
\begin{align}
    \sqrt{K} \delta T_\parallel &= \frac{M r}{\Sigma^2 \Delta} \qty( a \frac{4 a r z \dv{r_\text{g}}{\lambda} \dv{z_\text{g}}{\lambda} + 2 (r^2 - a^2 z^2)P_r P_z}{\Sigma} - (K - a^2 z^2)(r^2 - a^2) ) \,, \\ 
    \sqrt{K} \delta \Phi_\parallel &= \frac{1}{2 \Sigma^2 \Delta (1-z^2)} \qty( - \frac{\qty(4 a r z \dv{r_\text{g}}{\lambda} \dv{z_\text{g}}{\lambda} + 2 (r^2 - a^2 z^2) P_r P_z) (\Delta - a^2(1-z^2))}{\Sigma} ) + a \qty( 2 z^2 (K + r^2) \Delta - r (1-z^2)(K - a^2 z^2)\Delta' ) \,, \\
    \sqrt{K} \delta R_\parallel &= \frac{4 P_r \qty( a^2 z r \dv{r_\text{g}}{\lambda} \dv{z_\text{g}}{\lambda} - r^2 R )}{\Sigma^2} + \frac{P_r \qty( 2 R + r R' )}{\Sigma} - 2 E R + r P_r \Delta' + 2 ((a^2-r^2)E - a J_z) \Delta \,, \\
    \sqrt{K} \delta Z_\parallel &=  \frac{4 a P_z \qty( r z \dv{r_\text{g}}{\lambda} \dv{z_\text{g}}{\lambda} - a^2 z^2 Z )}{\Sigma^2} + \frac{a P_z \qty( 2 Z + z Z' )}{\Sigma} - 2 E Z + 2 a (J_z (1-2 z^2) - a (1-z^2)) \,, 
\end{align}
\begin{align}
    \delta T_{\rm c} &= - \frac{2 M a \qty( 2 r \qty( (K (r^2 - a^2 z^2) - 2 a^2 r^2 z^2) \dv{r_\text{g}}{\lambda} \dv{z_\text{g}}{\lambda} - a r z (2K + r^2 - a^2 z^2) P_r P_z ) + z (K+r^2) (K - a^2 z^2) (r^2 - a^2) \Sigma )}{2 \sqrt{K (K+r^2)(K-a^2 z^2)} \Sigma^2 \Delta}  \,, \\
    \delta \Phi_{\rm c} &= \frac{ 2 (\Delta - a^2 (1-z^2)) \qty( (K (r^2 - a^2 z^2) - 2 a^2 r^2 z^2) \dv{r_\text{g}}{\lambda} \dv{z_\text{g}}{\lambda} - a r z (2K + r^2 - a^2 z^2) P_r P_z )}{2 \sqrt{K (K+r^2)(K-a^2 z^2)} \Sigma^3 \Delta (1-z^2)} \nonumber \\
    &\phantom{=} - \frac{ z \sqrt{(K+r^2) (K - a^2 z^2)} (r^2 - a^2) (2 r \Delta + a^2(1-z^2) \Delta')}{2 \sqrt{K} \Sigma^2 \Delta (1-z^2)} \,, \\
    \delta R_{\rm c} &= - \frac{2 a P_r \Xi (K (r^2 - a^2 z^2) - 2 a^2 r^2 z^2) \dv{r_\text{g}}{\lambda} \dv{z_\text{g}}{\lambda} }{\sqrt{K} \Sigma^2 (K - a^2 z^2)} + \frac{a z P_r \Xi R'}{\sqrt{K} \Sigma} + \frac{2 a r z R \Xi \qty( a \Sigma P_z - 2 (K - a^2 z^2) P_r )}{\sqrt{K} \Sigma (K - a^2 z^2)} \,, \\
    \delta Z_{\rm c} &= - \frac{2 P_z (K (r^2 - a^2 z^2) - 2 a^2 r^2 z^2) \dv{r_\text{g}}{\lambda} \dv{z_\text{g}}{\lambda} }{\sqrt{K} \Xi \Sigma^2 (K + r^2)} - \frac{r P_z Z'}{\sqrt{K} \Sigma \Xi} + \frac{2 a r z Z \qty( \Sigma P_r + 2 a (K + r^2) P_z )}{\sqrt{K} \Sigma \Xi (K + r^2 )} \,, \\
    \delta T_{\rm s} &= \frac{2 M r a \qty( -a z P_z \dv{r_\text{g}}{\lambda} + r P_r \dv{z_\text{g}}{\lambda} )}{\Sigma^2 \Delta \sqrt{(K + r^2)(K - a^2 z^2)}} \,, \\
    \delta \Phi_{\rm s} &= - \frac{ (\Delta - a^2 (1-z^2)) \qty( -a z P_z \dv{r_\text{g}}{\lambda} + r P_r \dv{z_\text{g}}{\lambda} )}{\Sigma^2 \Delta \sqrt{(K + r^2)(K - a^2 z^2)}} \,, \\
    \delta R_{\rm s} &= \frac{2 a R \qty( z \Xi \dv{r_\text{g}}{\lambda} + r \Xi^{-1} \dv{z_\text{g}}{\lambda} )}{\Sigma (K+r^2)} - \frac{2 a z P_r \Xi \dv{r_\text{g}}{\lambda} ((a^2-K^2) E - a J_z)}{(K+r^2) (K - a^2 z^2)} \,, \\
    \delta Z_{\rm s} &= \frac{2 Z \qty( z \Xi \dv{r_\text{g}}{\lambda} + r \Xi^{-1} \dv{z_\text{g}}{\lambda} )}{\Sigma (K - a^2 z^2)} - \frac{2 r P_z \Xi^{-1} \dv{z_\text{g}}{\lambda} ((a^2-K^2) E - a J_z)}{(K+r^2) (K - a^2 z^2)} \,.
\end{align}
\end{subequations}
Eqs.~\eqref{eq:eom_nonseparable} can be transformed to the new variables $\Tilde{x}^\mu = x^\mu + \delta x^\mu(x^\mu,\psi) + \order{s^2}$ with inverse transformation $x^\mu = \Tilde{x}^\mu - \delta x^\mu(\Tilde{x}^\mu,\psi) + \order{s^2}$ by transforming the $\tau$ derivatives on the left hand side as
\begin{equation}\label{eq:transf_dxdtau}
    \dv{x^\mu}{\tau} = \dv{\Tilde{x}^\mu}{\tau} - \pdv{\delta x^\mu}{x^\nu} \dv{x_\text{g}^\nu}{\tau} - \pdv{\delta x^\mu}{\psi} \dv{\psi}{\tau}
\end{equation}
and the $\Sigma$ functions on the left hand side with the geodesic functions on the right hand side as
\begin{equation}\label{eq:transf_fx}
    f(x^\mu) = f(\Tilde{x}^\mu) - \pdv{f}{x^\mu} \delta x^\mu \,.
\end{equation}
Because we work in the linear-in-spin regime, the linear corrections $\delta X(x^\mu)$ do not have to be transformed. By using Eqs.~\eqref{eq:transf_dxdtau} and \eqref{eq:transf_fx} in Eqs.~\eqref{eq:eom_nonseparable}, we obtain
\begin{align}
    \Tilde{\Sigma} \dv{\Tt}{\tau} &= T_r^{(E,J_z)}(\Tr) + T_z^E(\Tz) + a J_z - \Tilde{\Sigma} \partial_{\Tilde{y}} ((T_r^{(E,J_z)}(\Tr) + T_z^E(\Tz) + a J_z)/\Tilde{\Sigma}) \delta y + \delta T + \partial_{\Tilde{y}} \delta t \dv{y_\text{g}}{\tau} + \partial_\psi \delta t \dv{\psi}{\tau} \,, \\
    \qty(\mu \Tilde{\Sigma} \dv{\Tr}{\tau})^2 &= R_0 - \Tilde{\Sigma}^2 \partial_{\Tilde{y}} (R_0/\Tilde{\Sigma}^2) \delta y + \delta R + 2 \dv{r_0}{\lambda} \qty( \partial_{\Tilde{y}} \delta r \dv{y_0}{\lambda} + \partial_\psi \delta r \dv{\psi}{\lambda} ) \,,
\end{align}
where $y = r, z$, and similarly for $\Tilde{\phi}$ and $\Tilde{z}$. After calculating the derivatives and substituting the relations \eqref{eq:delta_eom}, the equations of motion for $\Tilde{x}^\mu$ can be expressed as \eqref{eq:eom_final}.

\subsection{Comparison with numerical trajectory}

\begin{figure*}
    \centering
    \includegraphics{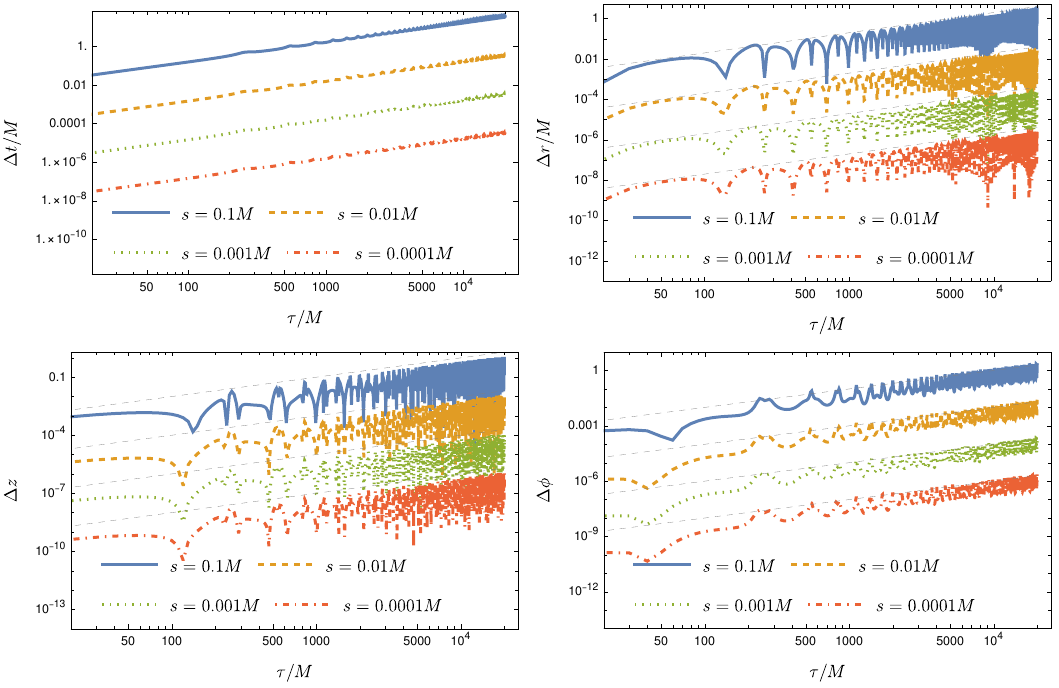}
    \caption{Difference between numerically and analytically calculated trajectory for an orbit with $a=0.95M$, $E = 0.960146$, $J_z = 2.48235 M$, $K = 8.91513 M^2$ and $s_\parallel = s/\sqrt{2} = s_\perp$. The difference decreases as $\order{s^2}$ which is consistent with the linearization and grows linearly with time which is caused by the $\order{s^2}$ differences in the frequencies.}
    \label{fig:error_numerical}
\end{figure*}

In this Subsection we describe the comparison of our analytical results with numerical integration of MPD equations \eqref{eq:MPD_full}. For the numerical integration we employed a Gauss-Runge-Kutta solver that was previously used in \cite{Skoupy:2023lih} to verify the trajectory of \citet{Drummond:2022b}. 

To compare the trajectories, we first need to find the initial conditions for the numerical solver. Because of the linearization, initial conditions calculated using the analytical results do not satisfy the nonlinearized constraints like the SSC or the normalization of four-velocity. Thus, we set the energy $E$, the angular momentum $J_z$, the initial coordinates $x^\mu_\text{i}$, $p^r_\text{i}$, $S^r_\text{i}$, $S^\theta_\text{i}$ and numerically calculated $p^z_\text{i}$, $S^t_\text{i}$ and $S^\phi_\text{i}$ from the constraints.

Then we numerically calculated the trajectory at equidistant points in $\tau$ and compared the values with the values obtained from analytical formulas \eqref{eq:result}. To find the values of $\Tl$ corresponding to the grid in $\tau$, we numerically inverted Eq.~\eqref{eq:result_tau}. The geodesic part of the trajectory was calculated using the \texttt{KerrGeodesics} package of the Black Hole Perturbation Toolkit \cite{BHPToolkit,KerrGeodesics}.

Figure \ref{fig:error_numerical} shows the difference $\Delta x^\mu = \abs{x^\mu_{\text{N}} - x^\mu_{\text{A}}}$ between the numerical trajectory $x^\mu_{\text{N}}$ and the analytical trajectory $x^\mu_{\text{A}}$ for different values of the spin $s$. We can see that as the spin $s$ decreases, the difference between the numerical and analytical results decreases as $\order{s^2}$, which is consistent with linearization, since the analytical solution is valid up to linear order (the numerical error of the solver is negligible here). 